\documentclass[11pt]{article}
\usepackage[margin=2 cm]{geometry}
\usepackage{comment}
\usepackage{graphicx}
\usepackage[nottoc]{tocbibind}
\usepackage{amsmath,amssymb,extarrows,mathtools,graphicx,subfigure,setspace}
\usepackage{cite}
\usepackage{braket}
\usepackage{epsfig}
\usepackage[section]{placeins}
\usepackage{slashed}
\usepackage{color}
\usepackage{caption}
\usepackage{amsmath}
\usepackage{hyperref}
\makeatother

\newcommand{\be}{\begin{equation}}
\newcommand{\bea}{\begin{eqnarray}}
\newcommand{\eea}{\end{eqnarray}}
\newcommand{\ba}{\begin{array}}
\newcommand{\ea}{\end{array}}
\newcommand{\ee}{\end{equation}}
\newcommand{\bes}{\begin{equation*}}
\newcommand{\beas}{\begin{eqnarray*}}
\newcommand{\eeas}{\end{eqnarray*}}
\newcommand{\bas}{\begin{array*}}
\newcommand{\eas}{\end{array*}}
\newcommand{\ees}{\end{equation*}}

\numberwithin{equation}{section}

\begin{document}

\onehalfspacing
\vfill
\begin{titlepage}
\vspace{10mm}

\begin{center}

\vspace*{10mm}
\vspace*{1mm}
{\Large  \textbf{ Phase transitions in Bergshoeff-Hohm-Townsend Massive Gravity }} 
 \vspace*{1cm}
 
{$\text{Mahdis Ghodrati}^{a,b} ,\text{ Ali Naseh}^b$}

\vspace*{8mm}
{ $^{a} $\textsl{Michigan Center for Theoretical Physics, Randall Laboratory of Physics\\ University of Michigan, Ann Arbor, MI 48109-1040, USA \\ \vspace*{4mm}
$^b $School of Particles and Accelerators,
Institute for Research in Fundamental Sciences (IPM) \\
P.O. Box 19395-5531, Tehran, Iran}} 
 \vspace*{1cm}

\textsl{E-mails: {\href{mailto:ghodrati@umich.edu}{ghodrati@umich.edu}}, {\href{mailto:naseh@ipm.ir}{naseh@ipm.ir}}    }
 \vspace*{2mm}

\vspace*{1.7cm}

\end{center}

\begin{abstract}

 We present the Hawking-Page phase diagrams in the Bergshoeff-Hohm-Townsend (BHT) massive gravity theory for different solutions, such as the
phase transitions between vacuum $\text{AdS}_3$ and BTZ black hole, warped $\text{AdS}_3$
and warped BTZ black hole in grand canonical and in non-local/quadratic
ensembles, Lifshitz black hole and the new hairy black hole solutions. We
observe that except for the black holes in quadratic ensemble, for other cases
in the non-chiral theory of BHT the phase diagrams are symmetric with respect to the direction of angular momentum, as we expected. We conclude that for presenting the phase diagrams of warped $\text{AdS}_3$ black holes, only the grand canonical ensemble should be used.

 \end{abstract}

\end{titlepage}

\tableofcontents


\section{Introduction}
In order to study quantum field theories with momentum dissipations, holographic ``massive gravity" theories could be exploited. There exist different three-dimensional massive gravity models with multiple geometrical solutions which have their own corresponding dual field theories. One of them is the ``Topological Massive Gravity'' (TMG) model, which is the Einstein-Hilbert action plus a Chern-Simons term that can break the parity of the pure Einstein theory. Recently in \cite{Detournay:2015ysa} the Hawking-Page phase transitions between $\text{AdS}_3$ and BTZ solutions, and also warped $\text{AdS}_3$ and warped BTZ black holes of TMG have been investigated and the Gibbs free energies, local and global stability regions and the phase diagrams of this theory have been presented.

There is yet another rich theory, the parity preserving Bergshoeff-Hohm-Townsend (BHT) or the ``New Massive Gravity'' (NMG) which has many interesting solutions which among the others are thermal warped $\text{AdS}_3$ and warped BTZ black holes. Similar to \cite{Detournay:2015ysa}, the aim of this paper is to study Hawking-Page phase transitions between different solutions of NMG. Particularly, we study phase transitions between thermal $\text{AdS}_3$ and BTZ black holes, warped $\text{AdS}_3$ and warped BTZ black holes in two different thermodynamical ensembles and the phase transitions between Lifshitz and also new hairy black holes with their corresponding vacua.

One should note that warped $\text{AdS}_3$ geometry is in fact a deformed $\text{AdS}_3$ that preserves the $SL(2,R) \times U(1)$ subgroup of $SL(2,R) \times SL(2,R)$ isometry. The obtained space-times are called null, time-like or space-like warped $\text{AdS}_{3}$ (WAdS$_{3}$) based on the norm of $U(1)$ killing vectors. The time-like WAdS$_{3}$ is famously called $G\ddot{o}del$ solution \cite{Rooman:1998xf, Reboucas:1982hn}.  As the phase transitions from the thermal $\text{AdS}_3$ or WAdS$_{3}$ to BTZ or warped BTZ black hole is dual to confining/deconfining phase transitions in the dual field theory, these models could be used in studying QCD or two-dimensional condensed matter systems with dissipations.

The plan of this paper is as follows. First, in section \ref{sec:num1}, by finding the free energies, we discuss phase transitions between the vacuum $\text{AdS}_3$ and BTZ black hole solutions and then in section \ref{sec:AdS1}, for the quadratic/non-local ensemble we discuss the thermodynamics and also the local and global stability regions. In section \ref{sec:sec2}, we calculate the free energies of warped $\text{AdS}_3$ vacuum and warped BTZ black hole solutions. We then present the phase diagrams of these solutions. For the sake of comparison, in section \ref{grandensemble} we present the phase diagrams in grand-canonical ensemble.  Next, in section \ref{sec:hairy}, we discuss the free energies and phase transitions of Lifshitz and new hairy black hole solutions of NMG and finally we finish up with a conclusion in section \ref{sec:disc}.  

\section{The Bergshoeff-Hohm-Townsend Theory }\label{sec:num1}

The Bergshoeff-Hohm-Townsend (BHT) or the new massive gravity (NMG) is a higher-curvature extension of the Einstein-Hilbert action in three dimensions which is diffeomorphism and parity invariant. In the linearized level, it is equivalent to the unitary Pauli-Fierz action for a massive spin-2 field \cite{Bergshoeff:2009hq}.

The action of NMG is
\begin{gather}\label{eq:action}
S=\frac{1}{16 \pi G_N} \int d^3x \sqrt{-g} \Big[ R-2\Lambda+\frac{1}{m^2} \Big( R^{\mu\nu} R_{\mu\nu}-\frac{3}{8} R^2 \Big)  \Big],
\end{gather}

where $m$ is the mass parameter, $\Lambda$ is the cosmological constant and $G_N$ is the three-dimensional Newton constant. In the case of $m \to \infty$, the theory reduces to the Einstein gravity action and in the limit of $m \to 0$, it would be just a pure fourth-order gravity model.  

The equation of motion from the action could be derived as
\begin{gather}
R_{\mu \nu}-\frac{1}{2} R g_{\mu\nu} +\Lambda g_{\mu\nu}+\frac{1}{m^2}K_{\mu\nu}=0,
\end{gather}
with the explicit form of $K_{\mu\nu}$ as below \cite{Bergshoeff:2009hq},

\begin{gather}
K_{\mu\nu} = \nabla^{2}R_{\mu\nu}-\frac{1}{4}\left(
\nabla_{\mu}\nabla_{\nu}R+g_{\mu\nu}\nabla^{2}R\right)-4R^{\sigma}_{\mu}R_{\sigma\nu}
+\frac{9}{4}RR_{\mu\nu}+\frac{1}{2}g_{\mu\nu}\left
(3R^{\alpha\beta}R_{\alpha\beta}-\frac{13}{8}R^{2}\right).\nonumber\\
\end{gather}
An auxiliary action which make the variational principle well-defined would be
\begin{gather}
S_{\text{Aux}}=\frac{1}{16 \pi G_N} \int_\sigma d^3x \sqrt{-g} \left( f^{\mu \nu} (R_{\mu \nu}-\frac{1}{2} R g_{\mu \nu})-\frac{1}{4} m^2 (f_{\mu \nu} f^{\mu \nu}-f^2) \right),
\end{gather}
where $f_{\mu \nu}$ is a rank two symmetric tensor with the following form
\begin{gather}
f_{\mu \nu}=\frac{2}{m^2} (R_{\mu \nu}-\frac{1}{4} R g_{\mu \nu}).
\end{gather}

This theory admits many solutions such as vacuum $\text{AdS}_3$, warped $\text{AdS}_3$, BTZ black hole, asymptotic warped AdS black hole, Lifshitz, Schr$\ddot{\text{o}}$dinger and so on \cite{Bergshoeff:2009hq},  \cite{AyonBeato:2009nh}. By finding the \textit{on-shell} free energies, we construct the phase diagrams between several of these solutions. 

 \section{Phase transitions of $\text{AdS}_3$ solution} \label{sec:AdS1}

The vacuum $\text{AdS}_3$ solution of this theory is as below
\begin{gather}
ds_{\text{AdS}_3}^2=l^2 (d\rho^2-\cosh^2 \rho \ dt^2+\sinh^2 \rho \ d\phi^2 ),
\end{gather}
where \cite{Grumiller:2009sn}
\begin{gather}
1/l^2=2m^2(1 \pm \sqrt{1+\frac{\Lambda}{m^2}}),
\end{gather}
and the boundary where the dual CFT is defined is located at $\rho \to \infty$.

For this case we use the relation $G(T,\Omega) =TS[g_c]$ to find the Gibbs free energy, where $g_c$ is the Euclidean saddle and $\tau=\frac{1}{2\pi} (-\beta \Omega_E+i \frac{\beta}{l})$ is the modular parameter. We work in the regimes where the saddle-point approximation could be used.

First, we need to find the free energy of the vacuum solution. In \cite{Kraus:2005vz} and \cite{Kraus:2006wn}, the authors derived a general result for deriving the action of thermal $\mathrm{AdS}_3$ in any theory as
\begin{gather}\label{Kraus}
S_E \big(AdS(\tau ,\tilde{\tau} )\big)=\frac{i \pi}{12 l} (c \tau-\tilde{c} \tilde{\tau}).
\end{gather}
The modular transformed version of this equation would then give the thermal action of the BTZ black hole. By changing the boundary torus as $\tau \to -\frac{1}{\tau}$ and then by using the modular invariance, one would have 
\begin{gather}
ds^2_{\text{BTZ}}\left[-\frac{1}{\tau}\right]=ds^2_{\text{AdS}} [\tau],
\end{gather}
which would result in
\begin{gather}\label{KrausBH}
S_E \big(BTZ(\tau ,\tilde{\tau} )\big)=\frac{i \pi}{12 l} (\frac{c}{ \tau}- \frac{\tilde{c}}{ \tilde{\tau}}).
\end{gather}

In this equation the contribution of the quantum fluctuations of the massless field is neglected as for low temperatures (large $\beta$) they are suppressed.  

One should notice that this equation and its modular transformed version only work for the $\text{AdS}_3$ case and not particularly for the ``warped $\text{AdS}_3$" or ``asymptotically warped AdS black holes" as their boundary is not an exact AdS.

So now, by inserting the central charges of NMG \cite{Bergshoeff:2009aq,Liu:2009bk}
\begin{gather}
c_L=c_R=\frac{3l}{2G_N} \left( 1-\frac{1}{2m^2 l^2 }\right),
\end{gather}
and the modular parameter $\tau=\frac{1}{2\pi} (-\beta \Omega_E+i \frac{\beta}{l})$ in Eq \ref{Kraus},  we find entropy as below
\begin{gather}\label{eq:SS}
S_E=-\frac{1}{8 l T G_N} \Big( 1-\frac{1}{2m^2 l^2} \Big).
\end{gather}
Unlike the corresponding equation in the TMG case, this relation does not depend on the Euclidean angular velocity $\Omega_E$. This is because the NMG theory has chiral symmetry and the central charges are equal to each other that makes the terms containing $\Omega$ to vanish. Using \ref{eq:SS} we find the Gibbs free energy as
\begin{gather}\label{eq:GADS}
G_{AdS}(T, \Omega)=-\frac{1}{8 l G_N} \Big( 1-\frac{1}{2m^2 l^2} \Big).
\end{gather}
Just by considering this equation one can see that the stability condition of the vacuum $\text{AdS}_3$ in NMG is $m^2 l^2 >\frac{1}{2}$ which is of course different from the Einstein theory. 

Additionally the NMG theory also admits a general rotating BTZ black hole solution which is in the following form,
\begin{gather} \label{eq:BTZ}
ds^2=l^2 \Big[ -\frac{(r^2-r_+^2)(r^2-r_-^2)}{r^2}dt^2+\frac{r^2}{(r^2-r_+^2)(r^2-r_-^2)} dr^2+r^2(d\phi+\frac{r_+ r_-}{r^2} dt)^2 \Big].
\end{gather}
The Hawking temperature of this black hole is \cite{Nam:2010dd}
\begin{gather}
T_H=\frac{\kappa}{2\pi} =\frac{1}{2\pi l} \frac{\partial_r N}{\sqrt{g_{rr}} } \Big |_{r=r_+}=\frac{r_+}{2\pi l}\Big(1-\frac{r_-^2}{r_+^2} \Big),
\end{gather}
the entropy is
\begin{gather}
S_{BH}=\frac{\pi^2 l}{3} c( T_L+T_R),
\end{gather}
and the angular velocity at the horizon is defined as \cite{Nam:2010dd}
\begin{gather}
\Omega=\frac{1}{l} N^\phi (r_+)=\frac{1}{l} \frac{r_-}{r_+} .
\end{gather}
The left and right temperatures are given by \cite{Maldacena:1998bw}
\begin{gather}
T_L=\frac{r_+ +r_-}{2\pi l}=\frac{T}{1- l \Omega}, \ \ \ \ \ \ \ \ \ \ \ \ \  T_R=\frac{r_+-r_-}{2\pi l}=\frac{T}{1+l \Omega},
\end{gather}
and the left and right energies can be defined as below
\begin{gather}
E_L \equiv \frac{\pi^2 l}{6} c_L T_L^2,  \ \ \ \ \ \ \ \ \ \  \ \  \ \ E_R\equiv \frac{\pi^2 l}{6} c_R T_R^2.
\end{gather}
These parameters are related to the mass and angular momentum as \cite{Nam:2010dd}
\begin{gather}
M=E_L+E_R, \ \ \ \ \ \ \ \ \ J=l(E_L-E_R).
\end{gather}
The horizons of the BTZ black hole are located at
\begin{gather}
r_+=\sqrt{2\Big( \frac{M \tilde{l}^2 }{4} +\frac{\tilde{l}}{4} \sqrt{M^2 \tilde{l}^2 -j^2}  \Big) }=\frac{2\pi l T}{1-\Omega^2 l^2}, \ \ \ \ \ r_-=\sqrt{2 \Big( \frac{M \tilde{l}^2 }{4} -\frac{\tilde{l}}{4} \sqrt{M^2 \tilde{l}^2 -j^2} \Big ) }=\frac{2\pi \Omega l^2 T}{1-\Omega^2 l^2}.
\end{gather}
The central charges of BTZ black hole in NMG is 
\begin{gather}
c_L=c_R=\frac{3 l}{2G_N} \Big( 1-\frac{1}{ 2 m^2 l^2 } \Big).
\end{gather}
For theory to be physical, the central charge and the mass of the BTZ black hole should be positive which again sets the condition $m^2 l^2 >\frac{1}{2}$.

These parameters satisfy the first law of thermodynamics
\begin{gather}
dM=T_H dS_{BH}+\Omega dJ,
\end{gather}
and its integral satisfy the Smarr relation \cite{Nam:2010dd},
\begin{gather}
M=\frac{1}{2}T_H S_{BH}+\Omega J.
\end{gather} 

Now one can read the Gibbs free energy from the following relation
\begin{gather}
G=M-T_H S_{BH} -\Omega  J.
\end{gather}
Using all the above equations, the Gibbs free energy of the BTZ in NMG would be
\begin{gather}\label{eq:GBTZ}
G_{BTZ} (T,\Omega)=-\frac{\pi ^2 T ^2\left(2 m^2 l^2-1 \right)}{4 G_N m^2l\left( 1 - l^2 \Omega ^2\right)}.
\end{gather}

This result can also be rederived by considering the modular invariance. Therefore, using the relation \ref{KrausBH} and $G(T, \Omega)=T S[g_c]$ again denotes the applicability of \ref{Kraus} for the $\text{AdS}_3$ solution of NMG. 

From the relation for the Gibbs free energy, one can see that for small rotations $\Omega$ and regardless of the size of black hole's event-horizon, the thermal stability condition for BTZ black hole is $m^2 l^2 > \frac{1}{2}$ \cite{Myung:2015pua}. If this condition is satisfied, the Hawking-Page phase transition can occur between the BTZ black hole and the thermal solution. For the case of $m^2 l^2 < \frac{1}{2}$  the fourth-order curvature term is dominant and an inverse Hawking-Page phase transition between the BTZ black hole and the ground state massless BTZ black hole would occur \cite{Myung:2015pua}. We now extend these results to the higher angular momentums.

\subsection{The stability conditions}

In order to study the local stability of BTZ black hole, we find the Hessain ($H$) of its free energy $G(T,\Omega)$ in the following form
\begin{gather}
H=\left(
\begin{array}{ll}
 \frac{\partial ^2G}{\partial T^2} & \frac{\partial ^2G}{\partial T \partial \Omega } \\ \\
 \frac{\partial ^2G}{\partial \Omega  \partial T} & \frac{\partial ^2G}{\partial \Omega ^2} \\
\end{array}
\right)= \left(
\begin{array}{ll}
 \frac{ \pi ^2 \left(2 m^2 l^2-1\right)}{2 G_N m^2 \left(\Omega ^2 l^2-1\right)} & \frac{\pi ^2 l^2 \left(1-2 l^2 m^2\right)  T \Omega }{G_N m^2 \left(\Omega ^2 l^2-1\right)^2} \\ \\
 \frac{\pi ^2 l^2 \left(1-2 m^2 l^2 \right)  T \Omega }{G_N m^2 \left(\Omega ^2 l^2-1\right)^2} & \frac{\pi ^2  T^2 \left(2 m^2 l^2-1 \right)  \left(l^2 +3 l^4 \Omega ^2\right)}{2 G_N m^2 \left(\Omega^2 l^2-1\right)^3} \\
\end{array}
\right)
\end{gather}

In the region where both of its eigenvalues are negative, the system is stable. By finding the eigenvalues of the above matrix and then by assuming $G_N=l=1$,  the stable region of the parameters would be found as $m^2 >1$ and $\Omega^2<1$ for any $T$ which is similar to the stability region of TMG in \cite{Detournay:2015ysa}. 

For studying the global stability one should find the difference of the free energies of AdS and BTZ which in our case is
\begin{gather}
\Delta G=G_{AdS}-G_{BTZ}=\frac{2m^2 l^2-1}{4 G_N m^2 l} \left( \frac{\pi^2 T^2}{1- l^2 \Omega^2} -\frac{1}{4l^2} \right).
\end{gather}

\subsection{Phase diagrams}
\begin{figure}[ht!]
\centering 
\begin{minipage}{.5\textwidth}
  \centering  
\includegraphics[width=.7\linewidth]{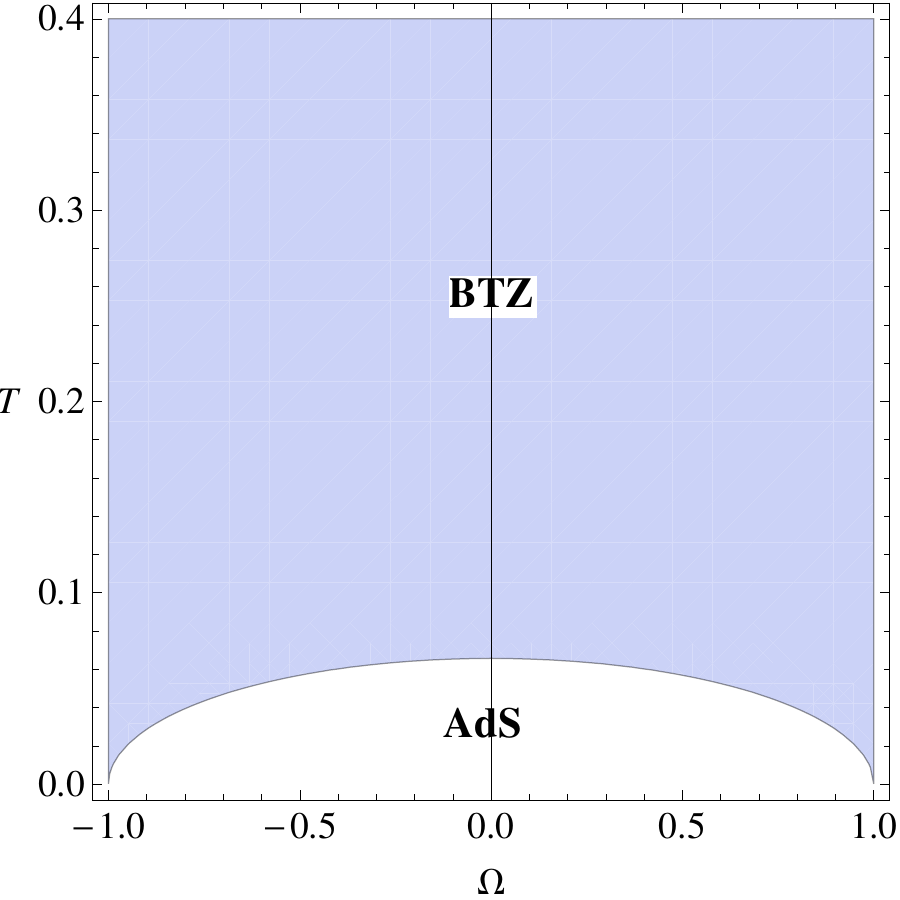} \caption{ \label{fig:p1}  $m=1.05$.} 
\end{minipage}%
\begin{minipage}{.5\textwidth}
\centering  
\includegraphics[width=.7\linewidth]{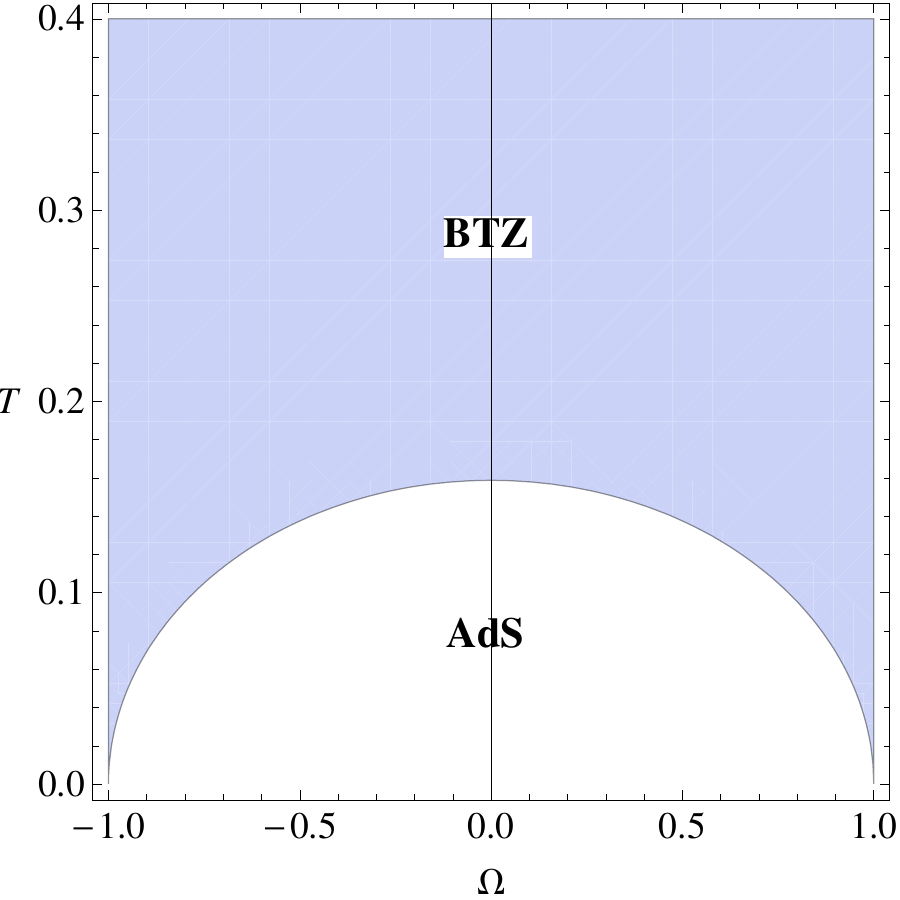}
\caption{\label{fig:p2}  $m=10$.}
\end{minipage}
\end{figure}

Based on thermodynamical relations, if $\Delta G>0$, the BTZ black hole is the dominant phase and if $\Delta G<0$, the phase of thermal $\text{AdS}_3$ would be dominant. By assuming $G_N=l=1$, we find the phase diagrams of Figures \ref{fig:p1} and \ref{fig:p2}. One can notice that these diagrams are symmetric since in NMG theory (unlike the TMG case) parity is conserved.

From the above diagrams one can also notice that by decreasing $m$ the effects of higher derivative correction terms in the action would increase. From this one can infer that the BTZ black hole in NMG would be formed in a lower temperature relative to pure Einstein-Hilbert theory and it is simply because the modes are massive here. On the other hand, increasing $m$ with a specific angular velocity would cause the phase transitions from $\text{AdS}_3$ to BTZ black hole occur at bigger temperatures.

\section{Phase transitions of warped $\text{AdS}_3$ solution in quadratic ensemble}\label{sec:sec2}

In order to study thermodynamics and the phase diagrams of black holes, one should first choose the thermodynamical ensemble. In \cite{Detournay:2012pc}, the authors introduced two different thermodynamical ensembles, the ``grand canonical" and the ``quadratic/non-local" ensembles. By specifically choosing the coordinate system to write the WAdS black hole (since the coordinate transformations are charge dependent), the ensemble of the black hole would be determined.

In this section, in the quadratic ensemble, we introduce the thermal $\text{WAdS}_3$ and Warped BTZ black hole solutions in the time-like and space-like coordinates and then we present the phase diagrams.

\subsection{G\"{o}del space-time}

The true vacuum of $\text{WAdS}_3$ black hole is the time-like $\text{WAdS}_3$ (the three-dimensional G\"{o}del spacetime) \cite{Donnay:2015iia,Banados:2005da}.  This metric is a constant curvature Lorentzian manifold with $U(1) \times SL(2, \mathbb{R})$ isometry group where the $U(1)$ factor is generated by a time-like Killing vector. 
Its metric is given by
\begin{gather}\label{eq:Godel}
ds^2=-dt^2-4 \omega r dt d\phi+\frac{\ell^2 dr^2}{(2r^2(\omega^2 \ell^2+1) +2\ell^2 r)}-\Big( \frac{2r^2}{ \ell^2} (\omega^2 \ell^2-1)-2r\Big) d\phi^2.
\end{gather}
In the special case of $\omega^2 \ell^2=1$, this metric corresponds to $\text{AdS}_3$. For this timelike solution we have
\begin{gather}\label{eq:LMG}
m^2=-\frac{(19\omega^2 \ell^2-2)}{2\ell^2},  \ \ \ \ \ \ \ \Lambda=-\frac{(11\omega^4 \ell^4+28 \omega^2 \ell^2-4)}{2\ell^2 (19 \omega^2 \ell^2-2)}.
\end{gather}

 Although this metric contains closed time-like curves (CTSs) and is unstable with respect to the quantum fluctuations, it has been studied to a great extent in the literature. In fact, this metric with large scale deficiencies surrounded by a more standard space-time can model some rotating objects which have physical applications in cosmology and condensed matter physics \cite{Rooman:1998xf}. Therefore one could think that constructing its phase diagrams would have some interesting applications.

\subsection{Space-like warped BTZ black hole}

In the quadratic/non-local ensemble and in the ADM form, the warped $\mathrm{AdS}_3$ black hole of NMG theory can be written as 
\begin{equation}
\begin{split} \label{eq:WBTZ}
\frac{ds^2}{l^2}= dt^2+\frac{dr^2 }{ (\nu^2+3)(r-r_+)(r-r_-) }+(2\nu r-\sqrt {r_+ r_- (\nu^2+3)} ) dt d\phi \\  
+\frac{r}{4} \Big[ 3(\nu^2-1)r +(\nu^2+3)(r_+ +r_-)-4\nu \sqrt {r_+ r_- (\nu^2+3)} \Big] d\phi^2.
\end{split}
\end{equation}
In this background, if $\nu^2=1$, the space is locally $\text{AdS}_3$, if $\nu^2 >1$, it is stretched and if $\nu^2<1$, it is a squashed deformation of $\text{AdS}_3$. For the space-like solution, the parameters would be
\begin{gather}\label{eq:mlBTZ}
m^2=-\frac{(20\nu^2-3)}{2 l^2}, \ \ \ \ \ \ \ \ \ \ \ \Lambda=-\frac{m^2 (4\nu^4-48\nu^2+9 ) }{(20\nu^2-3)^2}.
\end{gather}
If one employs the following relations
\begin{gather}
\omega=\frac{\nu}{l} , \ \ \ \ \ \ \ \ \   \  \omega^2 \ell^2+2=3 \ell^2/ l^2,
 \end{gather}
equation \ref{eq:LMG} can be re-derived.

Notice that ``$l$" is the radius of space-like $\text{AdS}_3$ and ``$\ell$" is the radius of the warped time-like $\text{AdS}_3$. 
Similar to the way we derived BTZ black hole by global identifications, we can also derive \ref{eq:WBTZ} from \ref{eq:Godel}.

In order to have a real $m$ and and a negative $\Lambda$ and therefore a physical solution, from \ref{eq:LMG} and \ref{eq:mlBTZ} the allowed range of $\nu$ and $\omega$ would be determined as
\begin{gather}
-\sqrt{\frac{2}{19}}<\omega \ell<\sqrt{\frac{2}{19}}, \ \ \ \ \ \ \ \ \ \ \ \  -\sqrt{\frac{3}{20} } <\nu < \sqrt{\frac{3}{20} }. 
\end{gather}

\subsection{The free energies and phase diagrams}\label{sub:free}
Now by using the thermodynamic quantities and conserved charges we calculate the free energies of both of these space-times and then we proceed by making the phase diagrams.

The conserved charges of timelike $\text{WAdS}_3$ in NMG for a \textit{``spinning defect"} have been calculated in \cite{Donnay:2015joa}. Using these relations one can take the limit of $\mu \to 0$ to find the mass and angular momentum as
\begin{gather}
\mathcal{M}=-\frac{4 \ell^2 \omega^2}{G_N(19 \ell^2 \omega^2-2)}, \ \ \ \ \ \ \ \ \ \ \ \ \ \ 
\mathcal{J}=-\frac{4 j \ell^4 \omega^3}{G_N(19\ell^2 \omega^2-2)}.
\end{gather}
The entropy and temperature of the time-like warped $\text{AdS}_3$ are zero. So the Gibbs free energy would be
\begin{gather}
G_\text{spinning defect }=\mathcal{M}-\Omega \mathcal{J}=\frac{4 \ell^2 \omega^2 \big((\mu-1)+\Omega j \ell^2 \omega \big)}{G_N(19\ell^2 \omega^2-2)}.
\end{gather}
Taking the limit of zero defect we get the following relation
\begin{gather}\label{eq:G1}
G_{\text{timelike WAdS}}=-\frac{4 \ell^2 \omega^2}{G_N (19 \ell^2 \omega^2 -2)}=-\frac{4\nu^2}{G_N (20\nu^2-3)}.
\end{gather}

Now by calculating the conserved charges we can calculate the thermodynamic properties and the Gibbs free energies of the black holes with asymptotic warped $\text{AdS}_3$ geometry.

 The thermodynamical quantities are \cite{Nam:2010dd}, \cite{Donnay:2015iia}
\begin{gather}\label{Tomega}
T_H=\frac{\nu^2+3}{8\pi \nu l} \bigg( \frac{r_+-r_-}{r_+- \sqrt{ \frac{(\nu^2+3)r_+ r_-}{4 \nu^2} } } \bigg), \ \  \ \ \ \Omega=\frac{1}{\nu l} \bigg( \frac{1}{r_+- \sqrt{ \frac{( \nu^2+3)r_+ r_-}{4\nu^2} } } \bigg),  
\end{gather}

\begin{gather}
T_L=\frac{(\nu^2+3)}{8 \pi l^2} (r_+ +r_- -\frac{1}{\nu} \sqrt{(\nu^2+3) r_- r_+} )  , \ \ \ \ \ T_R=\frac{(\nu^2+3)}{8 \pi l^2} (r_+ -r_-). 
\end{gather}
The conserved charges, mass and angular momentum are
\begin{gather}\label{Mass}
\mathcal{M}=Q_{\partial _t}= \frac{\nu (\nu^2+3) }{G_N l (20 \nu^2-3) } \Big( (r_-+r_+)\nu -\sqrt{r_+ r_-(\nu^2+3) } \Big) , 
\end{gather}
\begin{gather}\label{angular}
\mathcal{J} = Q_{\partial_{\varphi} }= \frac{\nu (\nu^2+3) }{4 G_N l ( 20\nu^2-3)} \Big( (5\nu^2+3) r_+ r_- -2\nu \sqrt{r_+ r_-(\nu^2+3) }  (r_+ +r_-) \Big) .
\end{gather}
Also the condition for the existence of black hole would be
\begin{gather}
\mathcal{J}  \le \frac{G_N l (20\nu^2-3)}{4\nu (\nu^2+3) } \mathcal{M}^2,  \ \ \ \ \ \ \ \ \ \mathcal{M} \ge 0.
\end{gather}
One can see that these inequalities specifically do not put any new constraint on $\nu$. In addition, the entropy of warped BTZ black hole in NMG is 
\begin{gather}\label{eq:entropy}
S_{BH}=\frac{8 \pi \nu^3}{(20\nu^2-3) G_N } (r_+-\frac{1}{2\nu} \sqrt{(\nu^2+3) r_+ r_- } )  .
\end{gather}

These thermodynamical quantities follow the first law of thermodynamics and their integrals follow the Smarr-like relation as below
\begin{gather}
M=T_H S_{BH}+2 \Omega J.
\end{gather}
The central charge of warped BTZ black hole is \cite{Donnay:2015iia} 
\begin{gather}\label{eq:centralcharge}
c=-\frac{96 l \nu^3 }{G_N (20\nu^4+57\nu^2-9) }.
\end{gather}

\begin{figure}[ht!]
\centering 
\begin{minipage}{.5\textwidth}
  \centering
\includegraphics[width=.7\linewidth]{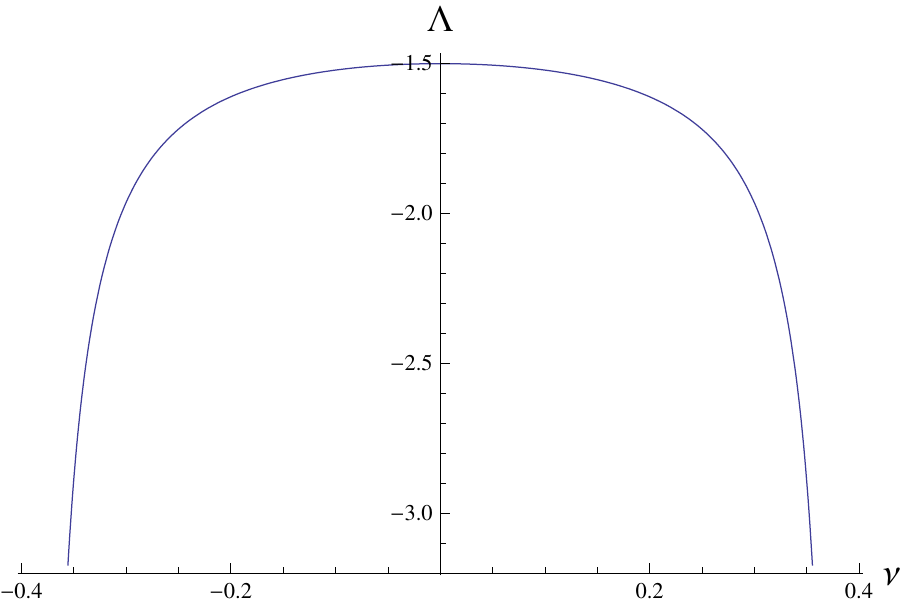} \caption{  \label{fig:c1} The plot of cosmological constant,  $\Lambda$ vs. $ -\sqrt{\frac{3}{20}}<\nu< \sqrt{\frac{3}{20}} $.} 
\end{minipage}%
\begin{minipage}{.5\textwidth}
\centering
\includegraphics[width=.7\linewidth]{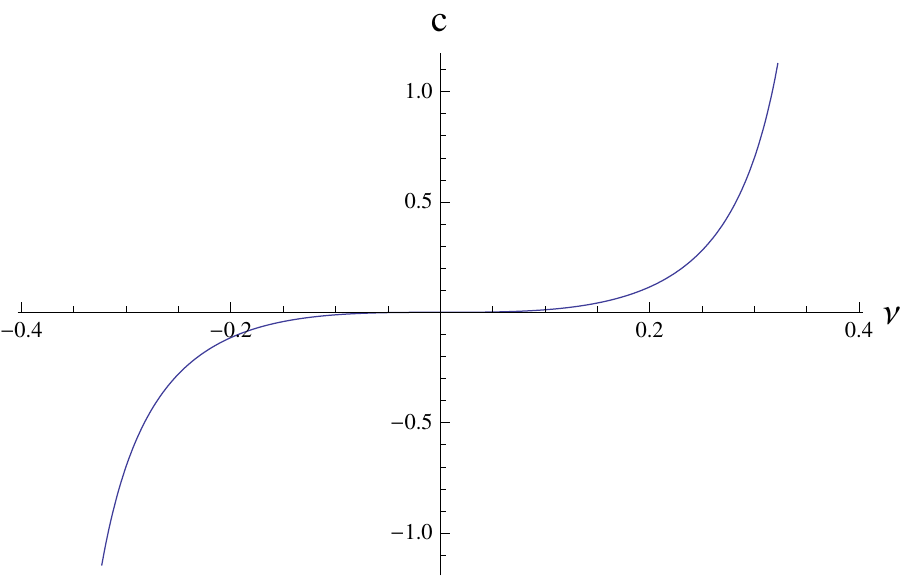} 
\caption{ \label{fig:c22}  The central charge of NMG vs. $\nu$.}
\end{minipage}
\end{figure}

One can now study the behavior of cosmological constant $\Lambda$ and also the central charge $c$ with respect to the warping parameter $\nu$ as in Figures \ref{fig:c1} and \ref{fig:c22}. Note that in the region where there is no CTSs, the change of central charge is a monotonically increasing function of $\nu$. Also at $\nu=0$ or $\nu\to \pm \infty$, the central charge is zero which indicates that for infinitely squashed or stretched space time the Casimir energy would vanishe. Note also that the central charge  diverges at  $\nu= \pm  \sqrt{\frac{3}{20}}\sim 0.387$. 

For characterizing a physical theory the central charge should be positive. If we assume $G_N=l =1$, then the constraint on $\nu$ would be $0<\nu <\sqrt{\frac{3}{20}}$.

If one defines 
\begin{gather}
K=1+\frac{8 l \pi  T (l \pi  T-\nu ) \pm 4 \pi  l T \sqrt{4 \pi ^2 l^2 T^2-8 l \pi  T \nu +\nu ^2+3}}{\nu ^2+3}\text{  },
\end{gather}
then
\begin{gather}\label{eq:radius}
r_+=\frac{1}{\Omega \nu l \Big(1-\frac{1}{2 \nu} \sqrt{K(\nu^2+3)}\Big)  }, \ \ \ \ \ \ \ \ \ \ \ \ \ \ \ \ \  r_-=K r_+.
\end{gather}
The minus sign in the relation for $K$ would be the correct sign as it leads to a smaller $r_-$ relative to $r_+$. Now in terms of $\Omega ,T$, $K$ and $\nu$ the Gibbs free energy would be
\begin{equation}
\begin{split}
G_{WBTZ}&=\frac{1 }{G_N  \Omega l (20 \nu ^2-3)}\Bigg[- 8 \pi  T \nu^2 +\frac{ (\nu ^2+3) }{l \Big(1-\frac{1}{2\nu}\sqrt{K(\nu^2+3) }  \Big) }\Bigg ( (K+1)\nu -\sqrt{K(\nu^2+3) }  \nonumber\\&
 -\frac{(5\nu^2+3) K-2\nu(K+1)\sqrt{K(\nu^2+3) }    }{4\nu l \big(1-\frac{1}{2\nu}\sqrt{K(\nu^2+3)} \big )  } \Bigg)\Bigg].
\end{split}
\end{equation}
One can see that this Gibbs free energy only depends on $\Omega$, $T$ and $\nu$. 

As mentioned earlier, the ensemble that this metric has been written and the phases have been found is the ``quadratic/non-local" ensemble \cite{Detournay:2012pc}. The partition function for this specific ensemble is \cite{Detournay:2012pc}
\begin{gather}
Z_{bh}= \text{Tr} \  \text{exp} \left [-\beta_R \left(L_0-\frac{P_0^2}{k} \right)-\beta_L \frac{P_0^2}{k} \right], 
\end{gather}
where  $\beta _{L,R} = T_{L,R}^{-1}$,  $L_0$ is the charge associated to the $SL(2,R)$ zero mode, $P_0$ is the $U(1)$ charge, and $c$ and $k$ are the central extensions of the Virasoro + Kac-Moody algebra.

In this ensemble, the corresponding algebra cannot be written as the variations of local currents, and its U(1) part is charge-dependent. Its algebra is a quadratic combination of Virasoro and Kac-Moody generators and it contains spectral flow and anomalous terms. Also all the currents of the algebra are right moving. 
We will show that although the microcanonical entropy does not depend on the choice of ensemble, but the Gibbs free energy and therefore the phase diagram does and therefore this ensemble, as it contains anomaly, might not be suitable for studying the phase transitions of the black hole solutions.

Now that we have found the free energies of the warped BTZ black hole and its vacuum, we can find the phase diagrams (temperature versus the angular velocity). Then we can compare them for different warping factors and therefore we would be able to study the effect of $\nu$ on the phase transitions.

\begin{figure}[ht!]
\centering 
\begin{minipage}{.5\textwidth}
  \centering
\includegraphics[width=.7\linewidth]{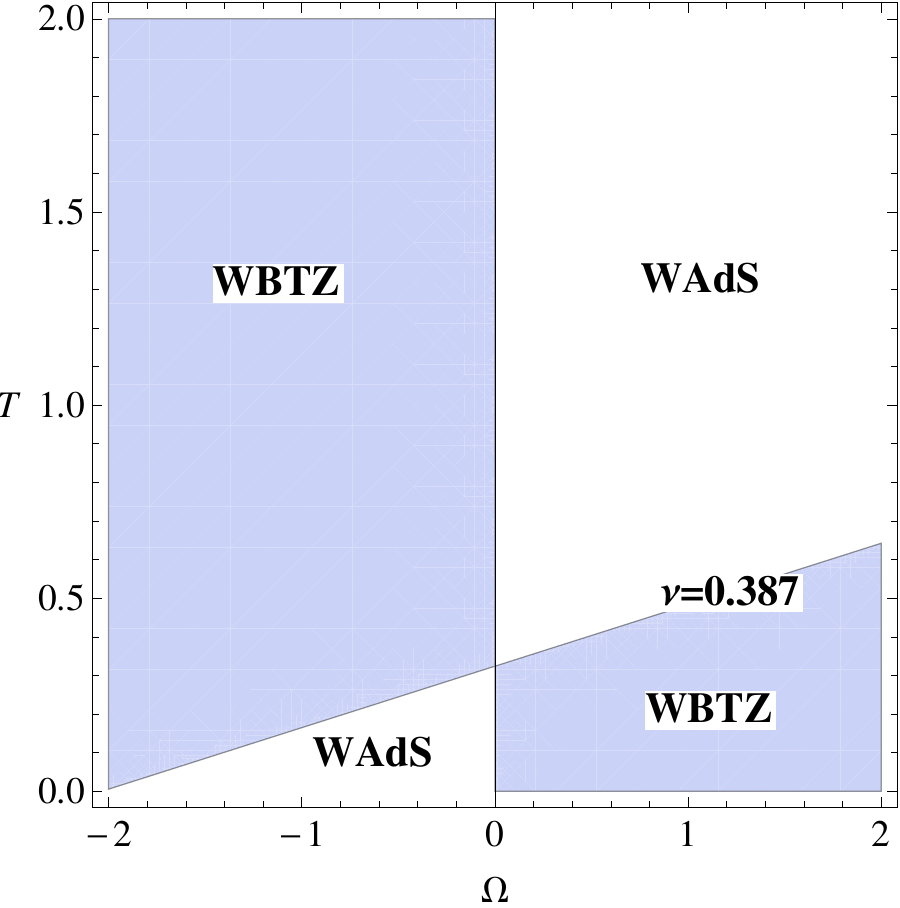} \caption{ \label{fig:nu1}   The phase diagram for $\nu=0.387$.} 
\end{minipage}%
\begin{minipage}{.5\textwidth}
\centering
\includegraphics[width=.7\linewidth]{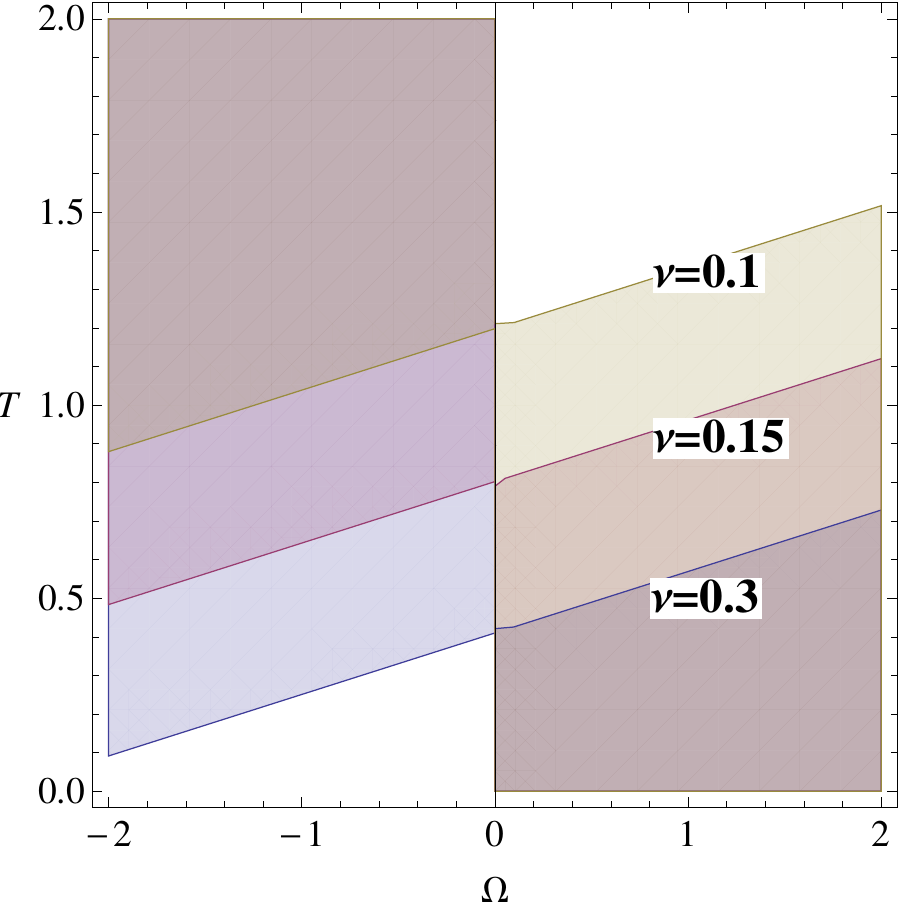}
\caption{  \label{fig:warp} The phase diagram for different $\nu$.}
\end{minipage}
\end{figure}

The phase diagram for $\nu=0.387$ is shown in Figure \ref{fig:nu1}. In the blue regions the warped BTZ black hole is the dominant phase and in the white regions the vacuum WAdS is dominant. If one increases $\nu$ till the relation $\nu \ge \sqrt{\frac{3}{20}}$ become satisfied, then these two phases would swapped with each other.
 
As one can see from Figures \ref{fig:nu1}, \ref{fig:warp}, in the quadratic/non-local ensemble (unlike the grand canonical ensemble), the diagrams of warped AdS solution are not symmetric although NMG is a parity preserving theory. The phase diagram for the positive and negative angular velocities are very different as for the warped geometry the free energy is an odd function of $\Omega$.

One can notice that if $\Omega>0$, at lower temperatures the phase of warped black hole is dominant and at higher temperatures the thermal warped $\text{AdS}_3$ case would be dominant.  In this case increasing temperature would trigger the reverse Hawking-Page transition and the black hole phase changes to the thermal warped $\text{AdS}_3$. So a bigger $\Omega$ makes the black hole solution the dominant phase while bigger $T$ makes the vacuum phase dominant.

The behavior of the phase diagram for different $\nu$ is also shown in Figure \ref{fig:warp}. For this diagram one can define a critical temperature $T_c$ where the tilted line crosses the $\Omega=0$ axis and then one can see that increasing $\nu$ would decrease this critical temperature. So for $\Omega <0$, increasing the warping factor $\nu$ makes the black hole phase dominant while in the case of $\Omega>0$, increasing $\nu$ would cause that the vacuum $\text{AdS}_3$ becomes the dominant phase.

\section{Phase diagram of warped $\text{AdS}_3$ solution in grand canonical ensemble} \label{grandensemble}

The WAdS$_3$ black hole in the grand canonical solution would be of the following form\\
\begin{gather}\label{metric}
g_{\mu \nu}=\left(
\begin{array}{ccc}
 -\frac{r^2}{l^2}-\frac{H^2 (-r^2-4 l J+8 l^2 M)^2}{4l^3(l M-J)}+8M & 0 & 4J-\frac{H^2(4 l J-r^2)(-r^2-4 l J+8 l^2 M) }{4l^2(l M-J)}  \\  
0 & \frac{1}{\frac{16 J^2}{r^2}+\frac{r^2}{l^2}-8M } & 0 \\ 
4J-\frac{H^2(4 l J-r^2)(-r^2-4 l J+8 l^2 M) }{4l^2(l M-J)} & 0 & r^2-\frac{H^2 (4 J l -r^2)^2 }{4 l (l M-J)} 
\end{array} \right).
\end{gather}

\begin{figure}[htb!]
\centering 
  \centering
\includegraphics[width=.35\linewidth]{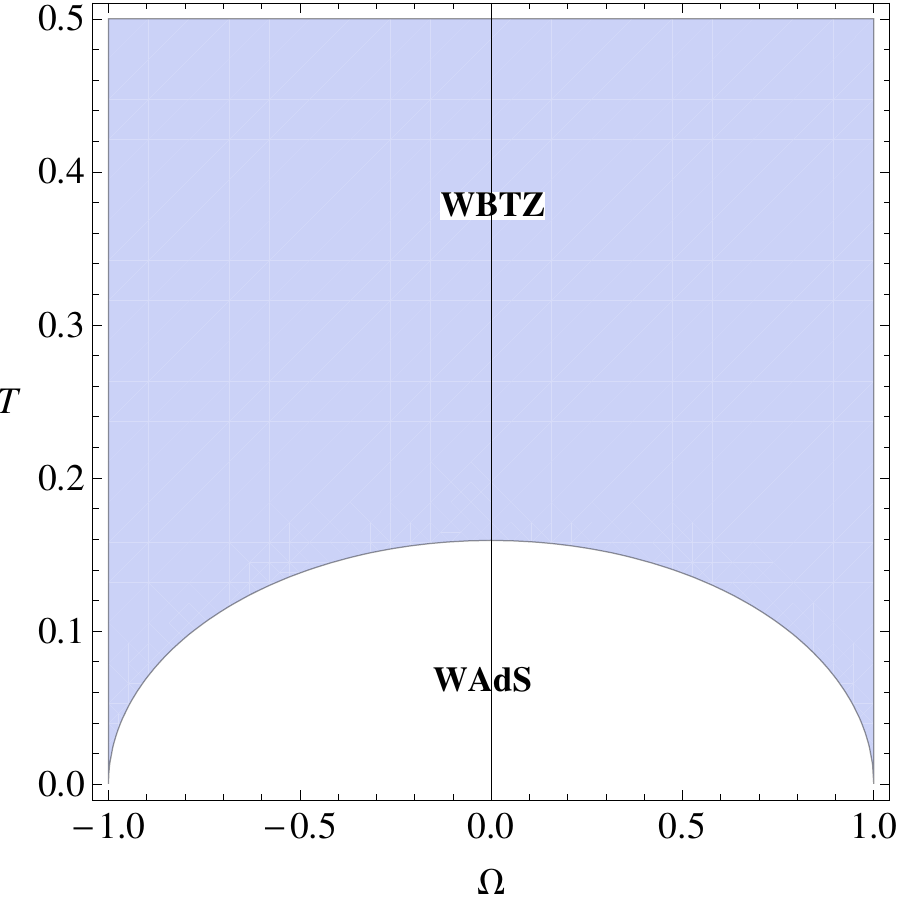} \caption{ \label{fig:grand}   Phase diagram for WAdS solution in grand canonical ensemble, $C=l=1$. } 
\end{figure}

The change of coordinate to go from the above metric to the form of \ref{eq:WBTZ} were derived in \cite{Detournay:2015ysa}. The theormodynamical ensemble of this black hole is the grand-canonical which its partition function is in the following form
\begin{gather}
Z \left (\beta, \theta \right)= \text{Tr}  e^{-\beta P_0+i \theta L_0},
\end{gather}

where $\beta$ here is the inverse temperature and $\theta$ is the angular potential.
The phase diagram for this specific ensemble was just recently presented  in \cite{Detournay:2016gao} which we brought here in Fig \ref{fig:grand} for the sake of comparison to the previous case.

Using the relations of conserved charges and entropy, one can derive the Gibbs free energy as below
\begin{gather}
G_{WBTZ} (T, \Omega)=\frac{-8 l^2 \pi^2 T^2 (1-2H^2)^{\frac{3}{2}} }{(17-42H^2)(1-l^2 \Omega^2) }.
\end{gather}

By setting $M=-\frac{1}{8}$ and $J=0$ the Gibbs free energy of vacuum could also be found. 
\subsection{local stability}
The Hessian matrix for the solution \ref{metric} would be 
\begin{gather}\label{hessian1}
\textbf{H}=\left(
\begin{array}{cc}
 \frac{16 l^2 \pi ^2\left(1-2 H^2\right)^{3/2} }{\left(17-42 H^2\right) \left(l^2 \Omega ^2-1\right)} & \frac{-32 l^4 \pi ^2 T \Omega \left(1-2 H^2\right)^{3/2} }{\left(17-42 H^2\right) \left(l^2 \Omega ^2-1\right)^2} \\ \\
 \frac{-32l^4 \pi ^2 T \Omega  \left(1-2 H^2\right)^{3/2} }{\left(17-42 H^2\right) \left(l^2 \Omega ^2-1\right)^2} & \frac{16\pi ^2 T^2 \left(1-2 H^2\right)^{3/2}\left(l^4+3 l^6 \Omega ^2\right)}{\left(17-42 H^2\right) \left(l^2 \Omega ^2-1\right)^3} \\
\end{array}
\right)
\end{gather}

For having a locally stable solution, all of the eigenvalues of the Hessian should be negative. If $\Omega=0$ then any $H^2 <\frac{17}{42}$ can make both of the eigenvalues of \ref{hessian1} negative and the solution stable.

One can notice that unlike the previous ensemble, in the grand canonical ensemble the diagrams of warped BTZ black hole solution are symmetric. This in fact is the result of parity preserving nature of BHT gravity. This could show us that the thermodynamical properties of warped black holes and therefore their Hawking-Page phase diagrams could only be meaningful in the grand canonical ensemble and not in any other ensembles. The significations of phase diagrams in other thermodynamical ensembles are not clear to us.

\section{Phase diagram of the hairy black hole }\label{sec:hairy}

Another interesting black hole solution in the New Massive Gravity is the ``New Hairy black hole'' which in this section we are going to study its Hawking-Page phase transitions.  This solution was first introduced in \cite{Oliva:2009ip} and later it was studied more in \cite{Nam:2010dd, Giribet:2009qz }.  

If the parameters of the actions \ref{eq:action} be $m^2=\Lambda=-\frac{1}{2 l^2}$, the hairy black hole could be a solution of BHT. The form of its metric is as follows
\begin{gather}
ds^2=-N F dt^2+\frac{dr^2}{F}+r^2 (d\phi+N^\phi dt)^2.
\end{gather}
The definition of the parameters in this metric is as below
\begin{equation}
\begin{split}
N=&\Big( 1+\frac{b l^2}{4H} \big( 1-\Xi^{\frac{1}{2}} \big) \Big)^2,\ \ \ \ \ \ \ \ \ \  
N^\phi=-\frac{a}{2r^2} (4G_N M-bH),\nonumber\\
F=&\frac{H^2}{r^2} \Big( \frac{H^2}{l^2} +\frac{b}{2} \Big(1+ \Xi^{\frac{1}{2}} \Big)H+\frac{b^2 l^2}{16} \Big( 1-\Xi^ {\frac{1}{2}} \Big)^2- 4G_N M\ \Xi^{\frac{1}{2}} \Big),\nonumber\\
H=& \Big( r^2-2G_N M l^2 \big (1- \Xi^{\frac{1}{2}} \big) -\frac{b^2 l^4}{16} \big(1-\Xi^{\frac{1}{2}} \big)^2 \Big)^{\frac{1}{2}},
\end{split}
\end{equation}
where $\Xi :=1-\frac{a^2}{l^2}$ and $-l \le a \le l $. 

There are two conserved charges for this black hole which are $M$ and $J=M a$ and a gravitational hair parameter $b$. Its thermodynamical quantities would be 
\begin{equation}
\begin{split}
\Omega= & \frac{1}{a} \Big( \Xi^{\frac{1}{2}}-1 \Big),\ \ \ \ \ \ \ \ \ \ \ \
T= \frac{1}{\pi l} \Xi^{\frac{1}{2}} \sqrt{2 G_N \Delta M \Big (1+\Xi^{\frac{1}{2}} \Big)^{-1}},\nonumber\\
S= & \pi l \sqrt{\frac{2}{G_N} \Delta M \Big( 1+\Xi^{\frac{1}{2}}  \Big) },\ \ \ \ \ \ \ \ \ \ \ \ \
\Delta M = M+\frac{b^2 l^2}{16 G_N}.
\end{split}
\end{equation}

Using all these quantities one can read the Gibbs free energy. We see that the region where the black hole can be locally stable for any $b$ is $\Omega^2 l^2 <1$. With these parameters we can simplify the relation of the Gibbs free energy as
\begin{gather}\label{GNBH}
G_{NBH} =\frac{l^2}{16G_N}\text{  }\left(\frac{16 \pi ^2 T^2 \left(5 l^2 \Omega ^2-1\right)}{\left(l^2 \Omega ^2-1\right)^2}-\frac{b^2 \left(3 l^2 \Omega ^2+1\right)}{l^2 \Omega ^2+1}\right). 
\end{gather}
Note that in addition to $\Omega$ and $T$ the Gibbs free energy also depends on the hair parameter $b$. One can also check this free energy would not vanish for any real $b$.

Now to study the local stability of the hairy black hole we calculate its Hessian as
\begin{gather}
\mathbf{H}=\left(
\begin{array}{ll}
 \frac{2 l^4 \pi ^2 \Omega ^2 \left(l^2 \Omega ^2+3\right)}{G_N \left(l^2 \Omega ^2-1\right)^2} & \ \ \ \  -\frac{4 l^4 \pi ^2 T \Omega  \left(5 l^2 \Omega ^2+3\right)}{G_N \left(l^2 \Omega ^2-1\right)^3} \\ \\
 -\frac{4 l^4 \pi ^2 T \Omega  \left(5 l^2 \Omega ^2+3\right)}{G_N \left(l^2 \Omega ^2-1\right)^3} & \ \ \ \ \frac{l^4 \left(b^2 \left(l^2 \Omega ^2-1\right)^4 \left(3 l^2 \Omega ^2-1\right)+24 \pi ^2 T^2 \left(l^2 \Omega ^2+1\right)^3 \left(1+5 l^2 \Omega ^2\left(2+l^2 \Omega ^2\right) \right)\right)}{4 G_N \left(l^2 \Omega ^2-1\right)^4 \left(l^2 \Omega ^2+1\right)^3} \\
\end{array}
\right).
\end{gather}

\begin{figure}[ht!]
\centering 
\begin{minipage}{.5\textwidth}
  \centering
\includegraphics[width=.7\linewidth]{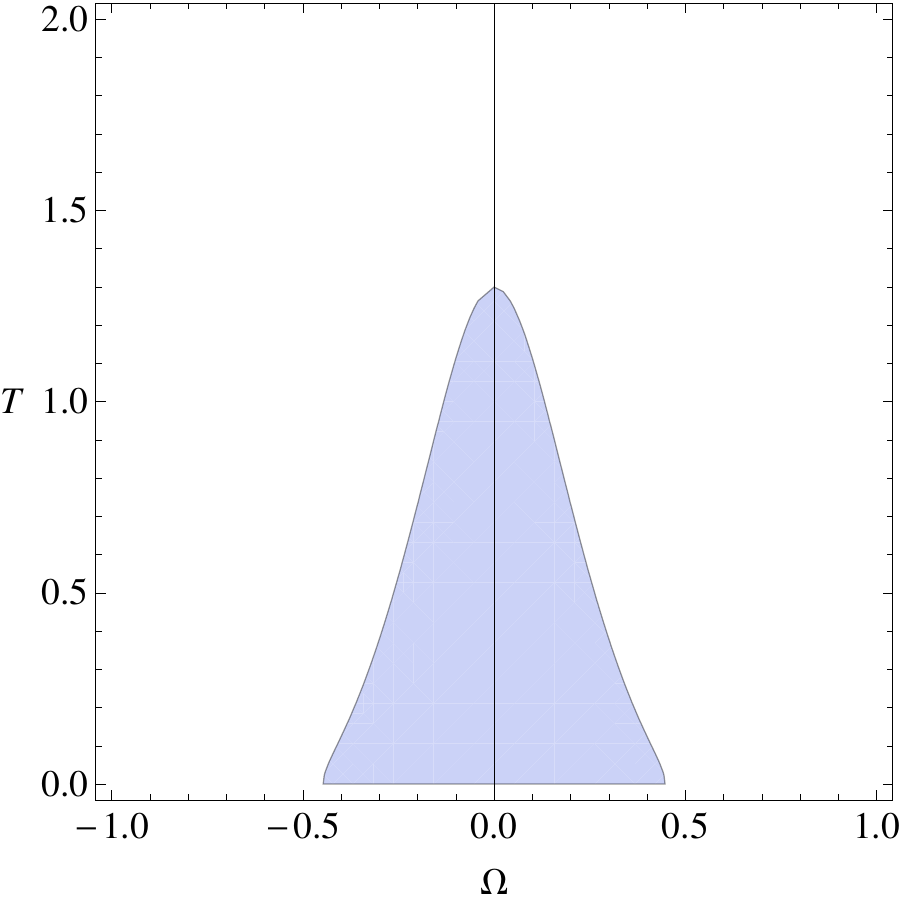} \caption{ \label{fig:localb1}  The local stable region for $b=20$. } 
\end{minipage}%
\begin{minipage}{.5\textwidth}
\centering
\includegraphics[width=.7\linewidth]{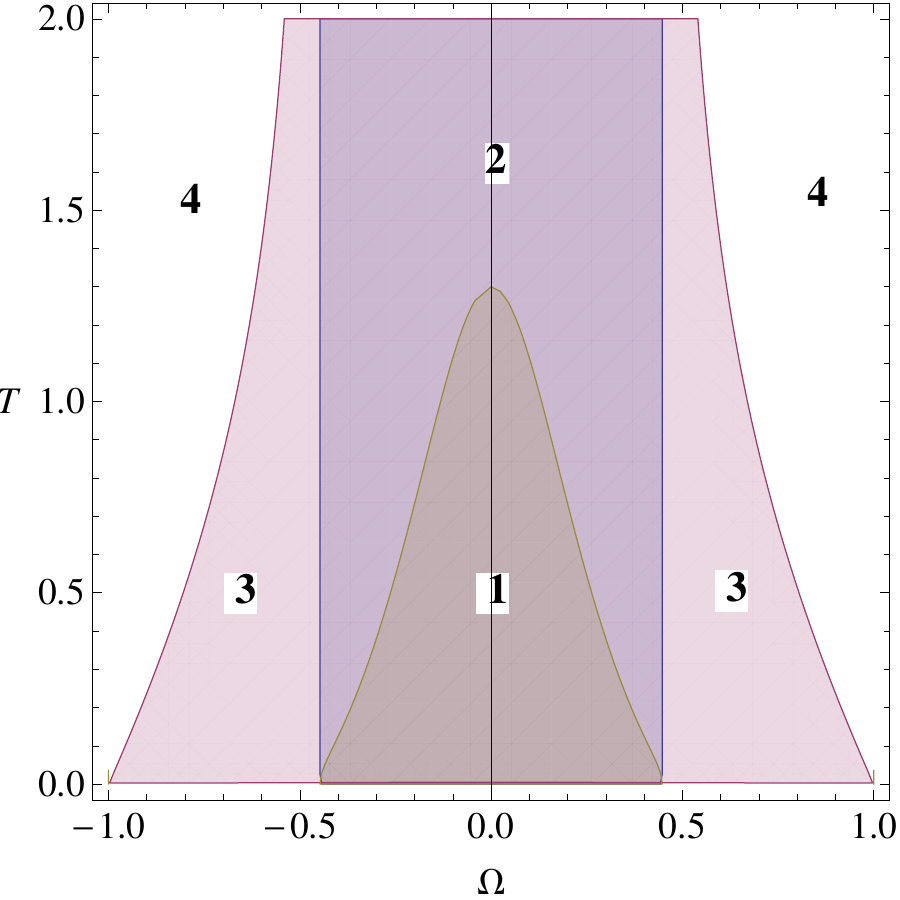}
\caption{ \label{fig:localb2}  The phase diagram for $b=20$.} 
\end{minipage}
\end{figure}

The region where both of the eigenvalues of the above matrix are negative and therefore the black hole is stable would depend on the hair parameter $b$. The phase diagram for the specific value of $b=20$ is shown in Figure \ref{fig:localb1}.

 One can check that for $G_N=l=1$ and for any $b$ the angular velocity should be in the range of $-0.5< \Omega <0.5$, so that the black hole solution can be locally stable.  Increasing $\Omega$ can make the black hole locally unstable. Also any condition for $T$ would depend on $b$. 

Increasing the hair parameter $b$ would make the region of local stability wider. So basically the hair parameter makes the system more stable and $\Omega $ makes it more unstable. In condensed matter systems, it would be interesting to investigate the dual interpretation of the hair parameter and  how physically it can increase the stability.

In Figure \ref{fig:localb2}, the phase diagram for the region of local stability, the vacuum $\text{AdS}_3$ solution and the ground state of the hairy black hole is shown. The region where $\Delta G_1=G_{\text{AdS}}-G_{\text{NBH}} >0$ is the union of regions 1, 2, 3. By comparing the Gibbs free energy of this black hole with the free energy of vacuum AdS, one can deduce that for any locally stable region and for any value of $b$, only the black hole phase would be present. So the only phase that is both locally and globally stable is the black hole solution. Outside of region 1 no phase would be locally stable. 

The case of $M=M_0=-\frac{b^2 l^2}{16 G_N}$ in this solution is the ground state which corresponds to an extremal case where both the left, right and the Hawking temperatures and also the entropy vanish \cite{Giribet:2009qz}. So its free energy would be
\begin{gather}
{G_0}_{\text{NBH}}=-\frac{b^2 l^2}{16 G_N} \left(3-\frac{2}{1+l^2 \Omega ^2}\right).
\end{gather}
The region where $\Delta G_2={G_0}_{\text{NBH}}-G_{\text{NBH}}>0$ is the union of 1 and 2. Again one can see that the black hole can be stable only if $-0.5 <\Omega <0.5$.

One can also make the phase diagrams of $M$ versus $J$ and study the effects of other physical parameters or conserved charges on the stabilities and on phase transitions. This could shed more light on other physical characteristics of the whole system of solutions.

\section{Conclusion}\label{sec:disc}

In this paper we calculated the free energies of the thermal $\text{AdS}_3$ and BTZ black holes, thermal warped $\text{AdS}_3$ and warped BTZ black holes in the quadratic/non-local and grand canonical ensembles and also the new hairy black hole in BHT gravity and then we plotted the Hawking-Page phase transition diagrams. We found symmetric diagrams for the solutions with $\text{AdS}_3$ geometry, warped AdS geometry in grand canonical ensemble and non-symmetric diagrams for the solution of warped $\text{AdS}_3$ geometry in quadratic/non-local ensemble. Since the theory of BHT is parity preserving, this could lead to the result that only the grand canonical ensemble should be used to present the physical phase diagrams. By using the phase diagrams, we also studied the effects of mass parameters, warping factor $\nu$ and the hair parameter on the stability of the solutions.

 \medskip

\bibliographystyle{JHEP}
\bibliography{Wref}
\end{document}